\documentclass[nature,superscriptaddress,reprint,floatfix]{revtex4-1}

\usepackage[plainpages=false, colorlinks=true, anchorcolor=blue, linkcolor=blue, citecolor=blue, bookmarks=false]{hyperref}
\usepackage{amsmath, amsthm, amssymb}
\usepackage{graphicx}

\usepackage{soul}
\usepackage{color}
\usepackage{dcolumn}
\usepackage{bm}
\usepackage[mathscr]{euscript}
\usepackage{mathrsfs}
\usepackage{amsbsy}

\newcommand{\DG}{\textcolor{black}}

\begin{document}

\title{Fast Non-Adiabatic Dynamics of Many-Body Quantum Systems}

\author{B. Larder}
\affiliation{Department of Physics, University of Oxford, Parks Road, Oxford OX1 3PU, UK}

\author{\DG{D.O. Gericke}}
\affiliation{Centre for Fusion, Space and Astrophysics, Department of Physics, University of Warwick, Coventry CV4 7AL, UK}

\author{S. Richardson}
\affiliation{Department of Physics, University of Oxford, Parks Road, Oxford OX1 3PU, UK}
\affiliation{AWE, Aldermaston, Reading, Berkshire RG7 4PR, UK}

\author{P. Mabey}
\affiliation{LULI -- CNRS, Ecole Polytechnique, CEA, Universit{\'e} Paris-Saclay, F-91128 Palaiseau Cedex, France}

\author{T. G. White}
\affiliation{Department of Physics, University of Nevada, Reno, Nevada 89557, USA}

\author{G. Gregori}
\email[Corresponding author: ]{gianluca.gregori@physics.ox.ac.uk}
\affiliation{Department of Physics, University of Oxford, Parks Road, Oxford OX1 3PU, UK}

\begin{abstract}
\end{abstract}

\maketitle


{\bf Modeling many-body quantum systems with strong interactions
is one of the core challenges of modern physics. A range of methods has
been developed to approach this task, each with its own idiosyncrasies, approximations, and realm of applicability\cite{Motta2017}. Perhaps the most successful and ubiquitous of these approaches is density functional theory (DFT)\cite{Hohenberg1964}. Its Kohn-Sham formulation \cite{Kohn1965} has been the basis for many fundamental physical insights, and it has been successfully applied to fields as diverse as quantum chemistry, condensed matter and dense plasmas \cite{White2013,Witte2017,Dharma-wardana2006,Bauernschmitt1996,Burke2012,Ziegler1991,Jones2015}. 
Despite the progress made by DFT and related schemes, however, there remain many problems that are intractable for existing methods. In particular, many approaches face a huge computational barrier when modeling large numbers of coupled electrons and ions at finite temperature 
\cite{Baczewski2016,Jones2015,Mabey2017}. Here, we address this
shortfall with a new approach to modeling many-body quantum systems. Based on the Bohmian trajectories formalism
\cite{Bohm1952,Bohm1953}, our new method treats the full particle
dynamics with a considerable increase in computational speed. As a result, we are able to perform
large-scale simulations of coupled electron-ion systems without employing the adiabatic Born-Oppenheimer approximation.}

\vspace{5pt}

Let us consider a many-particle electron-ion system at finite temperature. 
In calculating the dynamics of both the electrons and ions, we must account for the fact that the ions evolve multiple orders of magnitude more slowly than the electrons, as a result of their much higher masses. If we are interested in the long-time ionic dynamics (for example, the ion mode structure), we face a choice of how to deal with this timescale issue. We can either model the system on the time scale of the electrons -- non-adiabatically -- and incur a significant computational cost (a cost that is prohibitive in most simulation schemes); or we can model the system on the timescale of the ions -- adiabatically -- by treating the electrons as a static, instantaneously adjusting background. The latter approach is far cheaper computationally, but does not allow for a viable description of the interplay of ion and electron dynamics.

The method we propose here enables us to employ the former (non-adiabatic) approach, retaining the dynamic coupling between electrons and ions, by significantly reducing the simulation's computational demands. We achieve this by treating the system dynamics with linearized 
Bohmian trajectories. Having numerical properties similar to those of molecular dynamics for classical particles, our approach permits calculations previously out of reach: systems containing thousands of particles can be modeled for long (ionic) time periods, so that dynamic ion modes can be calculated without discounting electron dynamics.

To begin constructing our method, we consider Bohm's formulation of
quantum mechanics. We can imagine a classical $N$-body system as a 
``$N$-trajectory'' moving through \DG{3$N$-dimensional} configuration space. 
Bohm's theory treats an \DG{$N$-body} quantum system as an ensemble of \DG{these} 
classical $N$-trajectories, interacting through an additional $N$-body potential \DG{$V_B$. This Bohm potential} is a functional of the density of 
$N$-trajectories in configuration space, $\Phi$, and a function of the 
spatial position ${\bf x}$; \DG{that is, $V_B = V_B({\bf x}|\Phi)$}.
Provided \DG{matching initial conditions}, the Bohm ensemble \DG{of $N$-trajectories} reproduces the \DG{dynamics of the probability density -- as given
by the} Schr\"{o}dinger equation -- exactly \cite{Bohm1952} 
(see also Supplementary Information for more details). 

\DG{In its exact form, Bohm's formulation is as intractable as the $N$-body 
Schr\"{o}dinger equation, requiring simulation of a huge number of $N$-body interacting classical systems. However, we can} construct a fast computational
method \DG{solving} Bohm's theory \DG{by introducing a thermally averaged,
linearized Bohm potential.} The \DG{exact (but inaccessible)} calculation for a pure quantum state with many particles -- based on the theory above -- would require us to propagate an array of $N$-trajectories through time, at each step recalculating their density and thereby $V_B$ \DG{(see Fig.~\ref{fig:Linearization}a)}. Reliable calculations of density in $3N$-dimensional space would require a prohibitive number of $N$-trajectories, however, making this unfeasible. \DG{Here,} we
propose an alternative: as opposed to applying this theory to a pure quantum state, we consider propagating an array of thermally coupled $N$-trajectories in a similar manner, as a model of a system at finite temperature. Our core assumption is that the time evolution of a finite-temperature quantum system can be approximated by a similar procedure to that used for the pure state; we consider an array of $N$-trajectories, each coupled to a heat bath setting its temperature, evolving under a potential that is a functional of $N$-trajectory density in configuration space. This procedure can be seen as a trajectory-based analogue of the linearization of the Bohm potential over states in quantum hydrodynamics \cite{Manfredi2005}.

For simplicity, we focus initially on systems \DG{in thermal} equilibrium. We allow our thermal $N$-trajectories -- together modeling the probability density of our finite-temperature system -- to evolve in time under a linearized mean Bohm potential of the underlying pure states. In this \DG{linear} approximation, we replace the mean Bohm potential experienced, expressible as a sum over functionals of individual states, with a functional of a sum over individual states.

In addition to moving focus from a pure state to a more practically important finite-temperature state, the key feature of our approximation is that it dramatically reduces the computational expense required to simulate the system (as compared to the exact pure state case). Crucially, now that we are considering finite-temperature $N$-trajectories, we need only track a single $N$-trajectory through time, rather than an impractical number of them. This follows from two observations:

\begin{figure}[t]
\begin{centering}
\includegraphics[width=0.95\columnwidth]{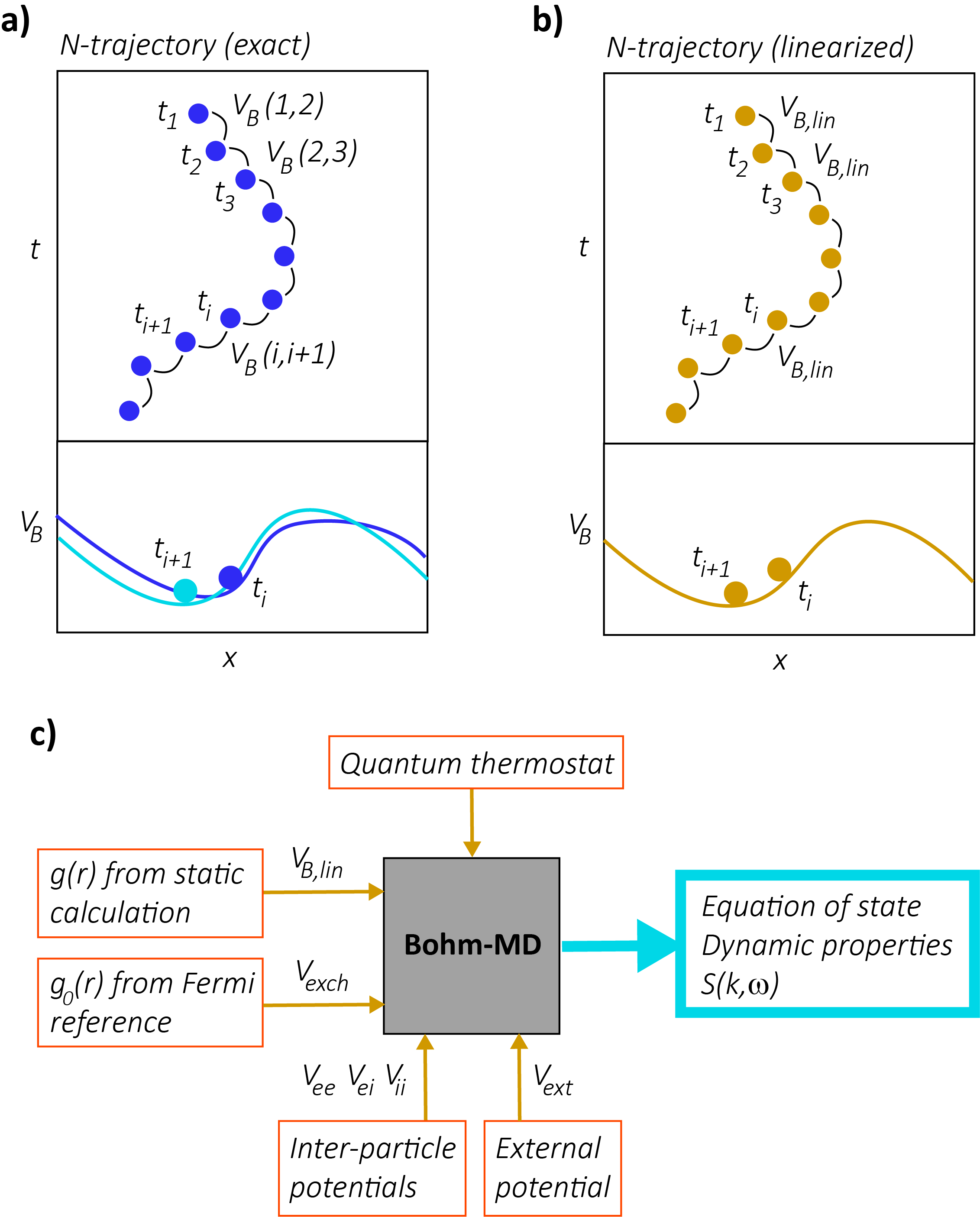}
\par\end{centering}
\caption{Schematic of the applied linearization approximation. a) The time evolution of an $N$-trajectory in an exact Bohmian representation of a pure quantum state (upper), and the Bohm potential, $V_{B}$, that it experiences (lower). $V_{B}$ is a functional of the density of $N$-trajectories in configuration space, $\Phi$. At each time step, all of the $N$-trajectories in the ensemble must be updated, $\Phi$ calculated, and the updated Bohm potential determined. b) The time evolution of an $N$-trajectory in our linearized Bohmian representation of a thermal state (upper), and the Bohm potential that it experiences (lower). We need only track a single $N$-trajectory: its coupling to a heat bath ensures its ergodicity, so that $\Phi$ becomes equal, in equilibrium, to its time-averaged density in configuration space. As a result, the $N$-trajectory evolves with a time-independent Bohm potential, generated self-consistently by its own time-integrated density. c) The block panel summarizes our scheme to determine the Bohmian dynamics and gives the sources for the different potentials needed as input.
\label{fig:Linearization}}
\end{figure}

{\it 1.~Properties of a classical system with correlation-dependent potentials can be determined self-consistently.} 
For an arbitrary system \DG{of $N$ well-localized particles}, the $N$-particle 
correlation function can be written as $g({\bf x})=P({\bf x})/P_{0}({\bf x})$,
where $P$ denotes the joint positional probability distribution of
the particles, \DG{and} $P_{0}$ is the distribution for a non-interacting
classical system with equal particle densities. Here, \DG{the variable}
${\bf x}$ is the set of particle positions, 
${\bf x}=\{{\bf x}_{1},{\bf x}_{2},...,{\bf x}_{N}\}$.
\DG{We may construct inter-particle potentials} that are functionals
of $g$: $V({\bf x})=V_{0}({\bf x})+V_{g}({\bf x}|g)$, where $V_{0}$ denotes pair interactions and external \DG{forces, and} $V_{g}$ is a \DG{contribution} that varies with $g$. The
equilibrium properties of a system in this potential can be found 
self-consistently. Starting from \DG{an initial guess for} $V({\bf x})$, we
can calculate $g({\bf x})$, through a Monte Carlo simulation or a similar
method \cite{Allen2017}. This value of \DG{the $N$-particle correlation 
function $g$} then gives rise to a new approximation for the potential.
Iterating this procedure allows for both $g$ and $V$ to be found. 

{\it 2.~Our linearized Bohmian system is equivalent to a classical system with 
correlation-dependent potentials.} 
Consider \DG{a number} of coupled thermal $N$-trajectories representing our 
linearized Bohmian system. Assuming that the system is in a temperature regime in which the particle motion is ergodic, we find that each 
$N$-trajectory has the same time-integrated correlations; \DG{that is,
each} has the same $g$. In the limit of infinite 
\DG{$N$-trajectories} in our ensemble, it follows that the configuration space density of $N$-trajectories $\Phi$ is exactly proportional to this common $g$. 
As a result, each $N$-trajectory moves in a common static potential (see Fig.~\ref{fig:Linearization}b),
\DG{and, as} this static potential is a functional of configuration space 
density, $\Phi$, it is equivalently just a functional of $g$.

\DG{When combining the results above,} our linearization approximation
 becomes a simple mapping:
\begin{equation}
V_{B}({\bf x}|\Phi) \mapsto V_g({\bf x}|g) 
     = -\frac{\lambda\hbar^{2}}{2\sqrt{g}} \,
        \sum_{i=1}^{N} \frac{\nabla_{i}^{2}}{m_{i}}\sqrt{g} \,,
        \label{eq:LinearizationApprox}
\end{equation}
where $m_i$ is \DG{the mass of particle $i$} and $\lambda$ is a linearization
factor \DG{that still needs to be} determined. As $g$ is common to all 
$N$-trajectories, our approximation scheme allows us to consider just a single $N$-body classical system (Fig.~\ref{fig:Linearization}b). The required simulation is thus amenable to (computationally cheap) classical Molecular
Dynamics  \cite{Allen2017}. The 
classical particle trajectories simulated then approximate the statistics of
the full quantum system. 

\DG{Before the scheme above can be implemented, we must overcome a final
fundamental} hurdle: the full correlation function $g$ appearing in
Eq.~(\ref{eq:LinearizationApprox}) is too complicated to be modeled directly
(\DG{similar to} the full $N$-body wave\-function). \DG{Therefore,} we seek
an approximate closure for this object in terms of lower-order correlations, which can be calculated accurately.
For this \DG{goal}, we employ \DG{the pair product} approximation, whereby the
$N$-body correlation is replaced by a product of pair correlations.
\DG{Further}, we generalize the dependence on $\lambda$ to a set
of $\lambda_{ij}$ to accommodate different particle species in the pair
interactions. 

\DG{We need an additional correction to our potentials to fulfil the spin 
statistics theorem,} as the Schr{\"o}dinger equation -- and thus Bohmian
mechanics -- does not \DG{incorporate} particle spin directly. 
\DG{In particular, this correction will generate a Fermi distribution for the electrons in thermal equilibrium.} \DG{Similar to} successful approaches
applied \DG{in} quantum hydrodynamics and classical map methods
\cite{Manfredi2005,Lado1967,Dharma-Wardana2012}, we introduce an additional
Pauli potential term. This term is constructed such that exchange effects are reproduced exactly for a reference electron gas system. 
We also employ \DG{pseudo-potentials, commonly applied}
in modern DFT calculations, \DG{to represent core electrons bound in deep
shells} of the ions by an effective ion potential \DG{seen by the valence
electrons}.

\DG{Finally, it} remains to set the linearization parameters 
$\lambda_{ij}$ to fully determine the system's Hamiltonian. In this work, we
take $\lambda_{ij}=1$ for the ion-ion and ion-electron terms. \DG{To
determine the electron-electron parameter $\lambda_{ee}$, we match static
ion correlations -- that is, pair distribution functions -- obtained by 
DFT calculations.} This matching can be carried out rigorously with 
a generalized form of inverse Monte Carlo (see Supplement Information).
 \DG{In this way, we determine 
$\lambda_{ee}$} with only \emph{static} \DG{information} of the system. 
\DG{Subsequent} \emph{dynamic} simulations can then be carried out without
any free parameters. 

\DG{We can now implement the Bohmian dynamics method with a Molecular Dynamics simulation inclusive of the potentials discussed above. Within the microcanonical ensemble, we could simply integrate the equation of motion for the particles.
However, we want to simulate systems with a given temperature which requires
the coupling of the system} to a heat bath. \DG{Here,} we employ a modified version of the Nos\'e-Hoover thermostat. \DG{Its standard form} is popular in 
classical Molecular Dynamics studies, achieving reliable thermodynamic
properties while having minimal impact on the dynamics \cite{Basconi2013}. We
use this standard version to control the ion temperature. \DG{The electrons, however, should relax to a Fermi distribution, which is not possible with this  classical form.} We have thus produced a modified version of the Nos\'e-Hoover 
thermostat for \DG{the dynamic electrons}. \DG{It creates an equilibrium 
distribution} equal to that of a non-interacting electron gas of the same 
density (see Supplementary Information for details and derivation). 
With this, we ensure the ions interact with an electron sub\-system \DG{with a corrected energy distribution.}

To demonstrate \DG{the strength of} our new method, we \DG{apply it to}
warm dense matter (WDM). With densities comparable to solids and temperatures
of a few electron volts, \DG{WDM combines the need for quantum simulations 
of degenerate electrons with the description of a strongly interacting ion 
component. These requirements make WDM an ideal testbed for quantum simulations. 
Further, as the matter in the mantle and core of large planets is in a WDM state
\cite{Guillot1999,Kerley1972}, and experiments towards inertial 
confinement fusion exhibit WDM states transiently} on the path to ignition
\cite{Glenzer2010,Hu2015}, simulations of WDM are of crucial importance in modern applications. 

\DG{Key dynamic properties} of the WDM state can be \DG{represented by} the 
dynamic structure factor (DSF) \cite{Hansen2013}. This quantity \DG{also} 
connects theory and experiment: \DG{probabilities for} diffraction and
inelastic scattering are \DG{directly proportional to the DSF 
\cite{GarciaSaiz2008,Glenzer2009,Fletcher2015}, allowing testing of}
WDM models. Here, we focus on the ion-ion DSF that is defined via
\begin{equation}
S({\bf k},\omega) = \frac{1}{2\pi N} \int \exp(i\omega t)
     \left\langle \rho({\bf k},t)\rho(-{\bf k},0)\right\rangle dt \,,
\label{eq:Direct_DSF}
\end{equation}
where $N$ is the total number of ions and $\rho({\bf k},t)$ is the spatial 
Fourier transform of the \DG{ion density}. \DG{In the following,} we assume
the WDM system to be isotropic and spatially uniform.
\DG{Accordingly,} the structure factor depends only on the magnitude of the
wavenumber, $k=|{\bf k}|$. While \DG{the main contribution to $S(k,\omega)$
is due to direct Coulomb interactions between the ions, the modifications
due to screening are} strongly affected by quantum behavior in the 
electron component.

Recent work has shown that \DG{predictions from standard DFT simulations for 
the ion-ion DSF are problematic due to the use of the Born-Oppenheimer
approximation \cite{Mabey2017}. Employing a Langevin thermostat, it has been
found that the dynamics of the electron-ion interaction may strongly change
the mode structure -- in particular, the strength of the diffusive mode.}
However, this approach requires a very simple, uniform frequency dependence
for electron-ion collisions, which may not prove realistic in practice. 
\DG{It also contains} an arbitrary parameter, the Langevin friction, and 
\DG{is of limited predictive power as a result.}

We demonstrate here that our new method of Bohmian dynamics, \DG{that retains
the dynamics of the electron-ion interaction, can overcome the shortcomings
of previous approaches without introducing free parameters.} \DG{The specific
case} we consider is compressed liquid aluminum with a density of 
$5.2$\,gcm$^{-3}$ and a temperature of $3.5$\,eV, \DG{which allows for direct
comparison with previous results.} Full details of the corresponding 
simulations and input parameters are given in the Supplementary Information.

\begin{figure}
\begin{centering}
\includegraphics[width=0.95\columnwidth]{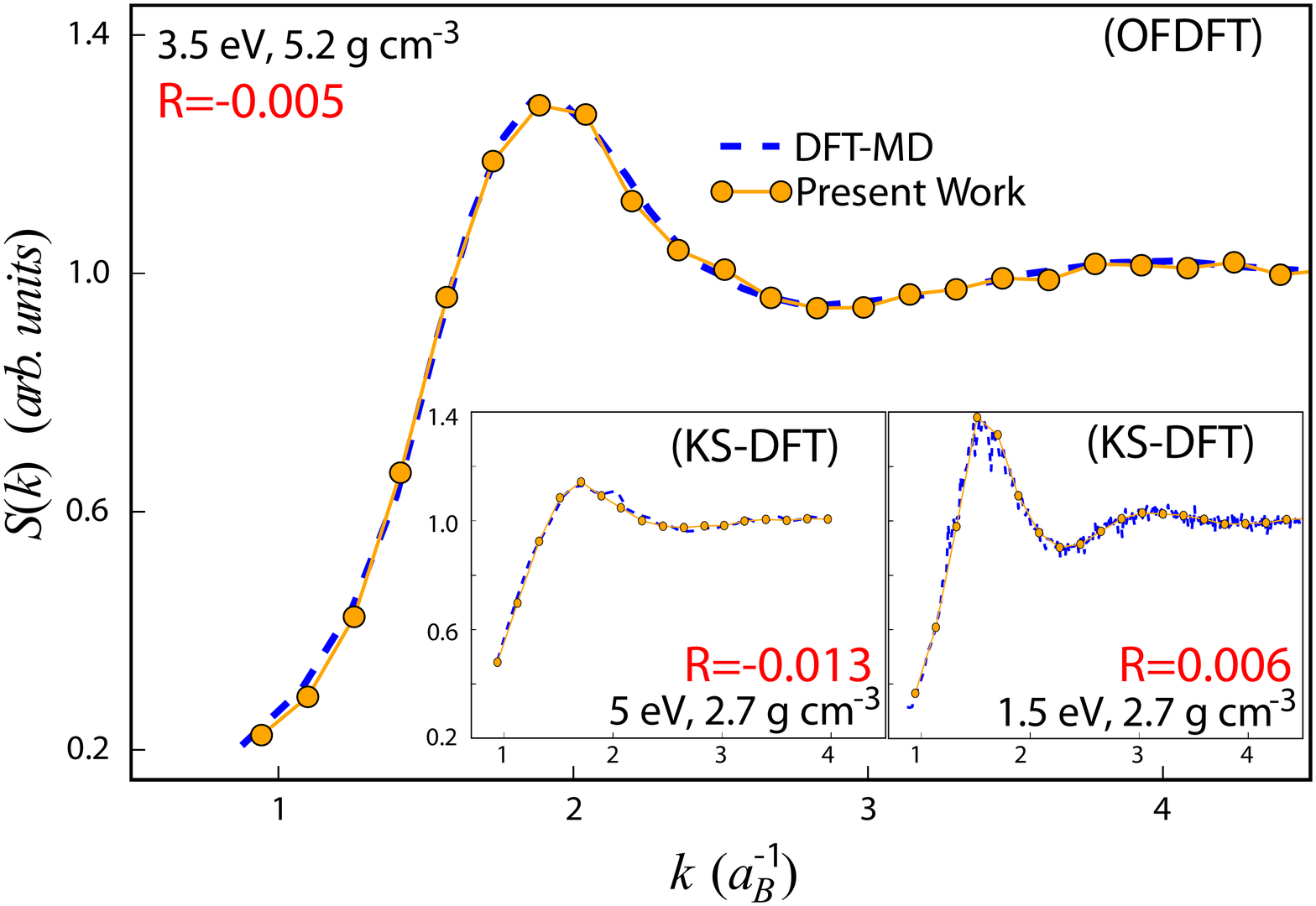}
\par\end{centering}

\caption{Static ion-ion structure factors \DG{for aluminum. The
         large graph compares our results from Bohmian dynamics
         with data obtained by orbital free DFT \cite{Mabey2017}
         for a density of $5.2\,$gcm$^{-3}$ and a temperature of
         $3.5\,$eV. The lower panels compare our results
         to data from full Kohn-Sham DFT simulations at solid density
         and two different temperatures. The excellent agreement of 
         the methods is also demonstrated by the very small 
         differences in pressure as quantified by the parameter $R$: 
         these values give the difference in ionic pressure between 
         the methods normalized to the difference of the 
         DFT pressures and the pressure of an ideal electron gas; 
         that is $R = (P_{Bohm}-P_{DFT})/(P_{DFT}-P_{0})$.}
         \label{fig:Static-ion-ion-correlation-comparison-with-dft}}
\end{figure}

The validity \DG{and accuracy of our implementation of Bohmian dynamics} 
is \DG{strongly} supported by the \DG{excellent} reproduction of static 
ion-ion correlations \DG{from DFT simulations}.
Fig.~\ref{fig:Static-ion-ion-correlation-comparison-with-dft} illustrates the
static \DG{ion-ion} structure factor obtained with the Bohmian trajectories
technique described above. \DG{The comparisons with orbital free DFT and 
the computationally more intensive Kohn-Sham DFT both yield agreement within
the statistical error of the simulations.} As discussed, \DG{this match was
achieved by a single parameter fit defining $\lambda_{ee}$. The 
different simulation techniques predict almost the same thermodynamics as 
shown by the small pressure difference.}

Fig.~\ref{fig:DSF-comparison,-calculated}a shows calculations of the 
\DG{fully} frequency-dependent DSF. \DG{One can} clearly notice the 
appearance of side peaks in the DSF that correspond to ion acoustic waves.
\DG{Their dispersion for smaller wavenumbers, and the corresponding sound
speed, are very sensitive to the interactions present in the system. Thus, they reflect the screening of ions by electrons as well as dynamic
electron-ion collisions. For larger wavenumbers $k$, these modes cease to
exist due to increased damping. The data also exhibit a diffusive mode 
around $\omega = 0$ although it is not as prominent as predicted in 
Ref.~\cite{Mabey2017}.}

\begin{figure}
\begin{centering}
\includegraphics[width=0.95\columnwidth]{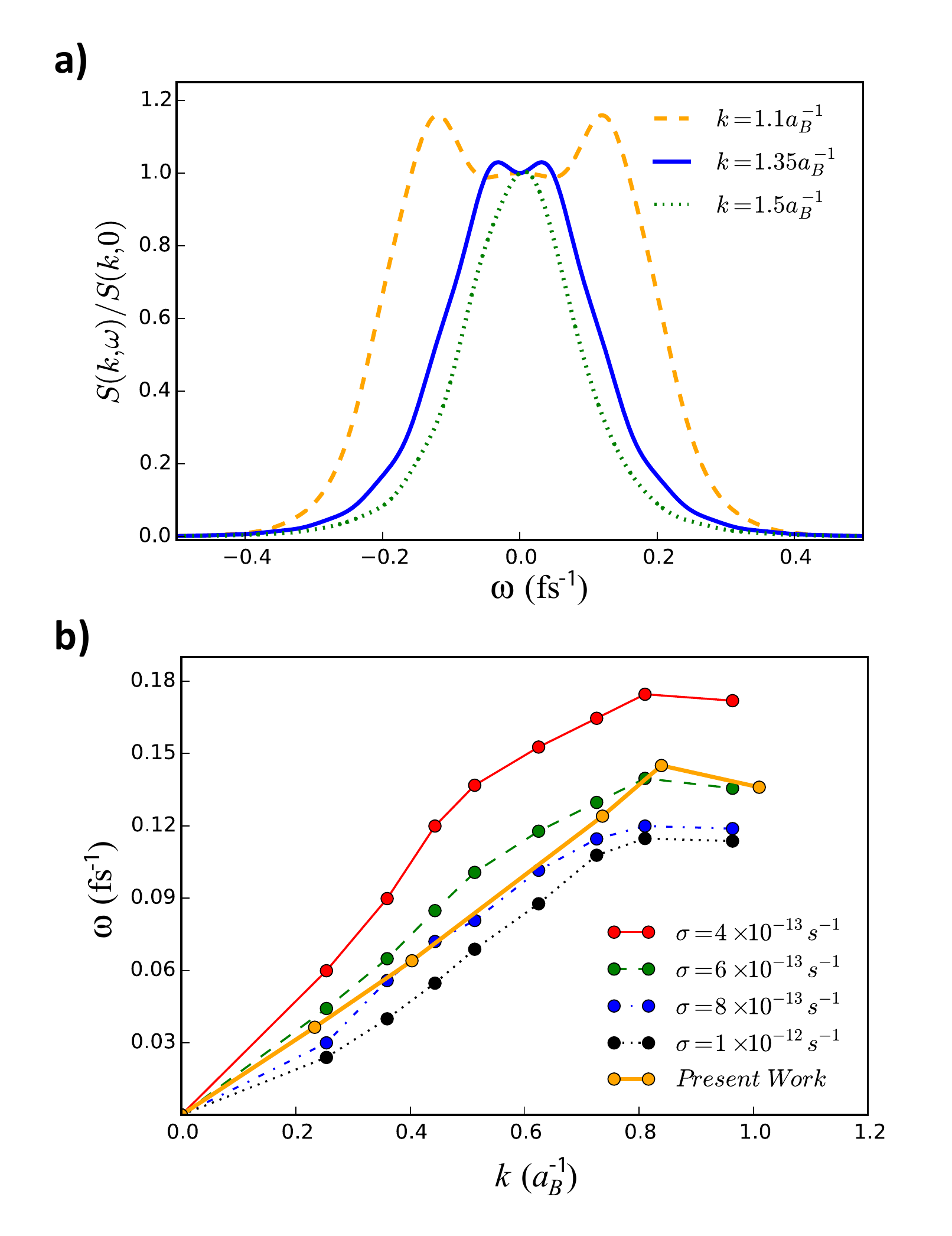}
\par\end{centering}
\caption{\DG{Results for the dynamic ion structure for aluminum
         at $3.5$\,eV and $5.2$\,gcm$^{-3}$. The upper panel
         shows the frequency-resolved DSF from the Bohmian
         dynamics. The lower panel provides a comparison of
         the dispersion relation of the ion acoustic modes 
         from our Bohmian approach with the data from the
         Langevin model of Ref.~\cite{Mabey2017}.}}
\label{fig:DSF-comparison,-calculated}
\end{figure}

\DG{The dispersion relation of the ion acoustic modes is displayed in 
Fig.~\ref{fig:DSF-comparison,-calculated}b, which also contains results from
the Langevin approach. The latter approach requires {\em ad hoc} friction
parameters which were chosen to cover the range between the classical and
quantum Born limits (see Ref.~\cite{Mabey2017}). The strength of the 
Bohmian approach lies in the fact that it does not require a free parameter, thereby allowing access to a self-consistent prediction of the sound speed. This
comparison may also be used to assess the quality of the friction
parameter applied in the Langevin approach. For the case considered, one finds that neither the
classical nor the weak coupling Born limit are applicable -- a finding which is typical of the WDM regime.} 

\DG{The upper results demonstrate the strength of the Bohmian approach in
modeling quantum systems with strong interactions and nonlinear ion dynamics. For static and thermodynamic properties, we obtain results 
in very close agreement with DFT simulations. In addition, while the standard
implementation of DFT involves the Born-Oppenheimer approximation, our 
Bohm approach can treat electrons and ions non-adiabatically, retaining the full coupling of the electron and ion dynamics. As a result, we can investigate
the changes of the ion modes due to dynamic electron-ion correlations which
are inaccessible to standard DFT. In contrast to a Langevin model, we have
no free parameters and can thus predict the strength of the electron drag
to the ion motion. Simulations based on time-dependent DFT \cite{Marques2004}
represent another way to avoid the Born-Oppenheimer approximation. However,
this method is numerically extremely expensive, drastically limiting
particle numbers and simulations times; at present, this limitation precludes results for the ion modes as presented here.}

The principal advantage of our approach is its relative numerical speed, which allows for the modeling of quantum systems with large numbers of particles.  For comparison,
the recent time-dependent DFT simulation of Ref.~\cite{Baczewski2016} models a system of 128 electrons for approximately 0.001 attoseconds per CPU-core and second. The comparative Bohmian dynamics system models 8 times as many electrons for approximately 20 attoseconds per CPU-core and second. This vast difference in computation time enables our method to access a new class of correlated quantum systems. Such  
calculations are relevant not just for WDM, but also address core problems in time-dependent chemical reactions, biological systems (e.g., protein folding), and radiation damage of materials \cite{protein,poccia,suga}.


\noindent
\subsection*{Acknowledgements}
\DG{This work has received} support from AWE plc., the Engineer\-ing and
Physical Sciences Research Council (grant numbers EP/M022331/1 and 
EP/N014472/1) and the Science and Technology Facilities Council of the 
United Kingdom. This material is partially based upon work supported by the U.S. Department of Energy, Office of Science, Office of Fusion Energy Science under Award Number DE-SC0019268.
\copyright British Crown Copyright 2018/AWE.

\subsection*{Author contributions}
This project was conceived by G.G.. The linearization theory, approximation scheme and numerical implementation were developed by B.L. with guidance from G.G. and D.O.G.. The paper was written by B.L., D.O.G. and G.G..
Supporting calculations and theory were provided by S.R., P.M. and T.G.W.. 

\subsection*{Data availability}
All data that support the findings of this study are available from
the authors upon request.

\subsection*{Author declaration}
The authors declare no competing financial interests.

\bibliographystyle{unsrt}

\end{document}


\title{Fast Non-Adiabatic Dynamics of Many-Body Quantum Systems: \\[0.5ex]
       Supplementary Information}

\author{B. Larder}
\affiliation{Department of Physics, University of Oxford, Parks Road, Oxford OX1 3PU, UK}

\author{\DG{D.O. Gericke}}
\affiliation{Centre for Fusion, Space and Astrophysics, Department of Physics, University of Warwick, Coventry CV4 7AL, UK}

\author{S. Richardson}
\affiliation{Department of Physics, University of Oxford, Parks Road, Oxford OX1 3PU, UK}
\affiliation{AWE, Aldermaston, Reading, Berkshire RG7 4PR, UK}

\author{P. Mabey}
\affiliation{LULI -- CNRS, Ecole Polytechnique, CEA, Universit{\'e} Paris-Saclay, F-91128 Palaiseau Cedex, France}

\author{T. G. White}
\affiliation{Department of Physics, University of Nevada, Reno, Nevada 89557, USA}

\author{G. Gregori}
\email[Corresponding author: ]{gianluca.gregori@physics.ox.ac.uk}
\affiliation{Department of Physics, University of Oxford, Parks Road, Oxford OX1 3PU, UK}

\maketitle

\subsection*{Bohm theory of quantum mechanics}
For a single particle, the Bohm formalism follows from writing
the wavefunction $\psi$ as a product of amplitude and phase information
\begin{equation}
\psi({\bf x},t)=R({\bf x},t){\rm e}^{i\phi({\bf x},t)/\hbar},
\end{equation}
\noindent where $\bf x$ is the position vector and $t$ the time. Using this form for $\psi$, the Schr\"{o}dinger equation yields
\begin{equation}
\frac{\partial R^{2}}{\partial t}+\nabla\cdot\left(R^{2}\frac{\nabla\phi}{m}\right) \,,
\label{eq:Continuity}
\end{equation}
\begin{equation}
\frac{\partial\phi}{\partial t}+V_{ext}+V_{B}+\frac{(\nabla\phi)^{2}}{2m}=0\,,
\label{eq:HamiltonJacobi}
\end{equation}
\noindent where $V_{ext}$ is the external potential, and we have introduced the Bohm potential
$V_{B}$, given by
\begin{equation}
V_{B}({\bf x},t)=-\frac{\hbar^{2}}{2m}\frac{\nabla^{2}R}{R} \,.
\label{eq:BohmPotential1d}
\end{equation}
Eq.~(\ref{eq:Continuity}) provides a description of probability continuity, while Eq.~(\ref{eq:HamiltonJacobi}) is of the form of a Hamilton-Jacobi equation, with the additional quantum potential given by (\ref{eq:BohmPotential1d}). Bohm's method exploits the form of these equations by treating the quantum system as an ensemble of classical systems. 

We can treat $R^{2}$ as representing the density of a set of classical systems in configuration space. By then allowing the classical systems to evolve in time according to the modified Hamilton Jacobi equation, Eq.~(\ref{eq:HamiltonJacobi}), we can obtain the time evolution of $R^{2}$, representing the time evolution of the probability density of the quantum system.

Generalizing this idea to many-body systems, $R^{2}$ represents the probability density of classical
$N$-body points in configuration space. These phase space points within the ensemble evolve via 
\begin{equation}
\frac{d{\bf x}}{dt} \equiv\ {\bf v} = \frac{\nabla\phi({\bf x},t)}{m}\,,
\label{eq:trajectory_velocity}
\end{equation}
where $\bf x$ and $\bf v$ label the positions and velocities in configuration-space, respectively. This relationship to the velocities allows for a quasi-classical description of trajectories, or a large ensemble of $N$-body points to be evolved concurrently. The statistical properties of the classical trajectories then reproduce the quantum mechanical results.

\subsection*{Correlation Closure}
We seek a closure approximation to the full correlation function $g$ appearing in Eq.~(1) in the main paper, in such a way that the Bohm potential can be constructed for the system. The
simplest, non-trivial approximation of this kind is a direct product representation in terms of pair correlation functions
\begin{equation}
g({\bf x}_{1},{\bf x}_{2},...,{\bf x}_{N})\simeq\prod_{j>i}g({\bf x}_{i},{\bf x}_{j}) \,,
\label{eq:KirkwoodClosure}
\end{equation}
where the pair correlation $g({\bf x}_{i},{\bf x}_{j})$ is the full correlation function with all other coordinates integrated out
\begin{equation}
g({\bf x}_{i},{\bf x}_{j})=\frac{\left(\prod_{k\neq i,j}\int d{\bf x}_{k}\right)g({\bf x}_{1},{\bf x}_{2},...,{\bf x}_{N})}{\Omega^{N-2}} \,,
\end{equation}.
Here, $\Omega$ is a normalization volume.

Combining Eqs.~(1) of the main paper and Eq.~(\ref{eq:KirkwoodClosure})
we arrive at an easily calculable expression for the Bohm potential
\begin{equation}
V_{B}({\bf x}|g)\simeq-\frac{\hbar^{2}}{2}\sum_{j>i}\frac{\lambda_{ij}(\frac{\nabla_{i}^{2}}{m_{i}}+\frac{\nabla_{j}^{2}}{m_{j\ }})\sqrt{g({\bf x}_{i}, {\bf x}_{j})}}{\sqrt{g({\bf x}_{i}, {\bf x}_{j})}} \;.
\label{eq:BohmFull}
\end{equation}
For computational efficiency, we have retained interactions up to pair terms only, and have generalized the dependence on $\lambda$ to a set of $\lambda_{ij}$ to accommodate different particle species. 

\subsection*{Fermi Statistical Corrections}\label{sec:fermi_stat_corr}
At the base level of the linearized Bohmian approach, we model the system within the Hartree approximation -- effects due to the electron exchange interaction are not accounted for. To rectify this, we introduce an additional potential term. By inverting the known pair correlations of the non-interacting electron gas using inverse Monte Carlo (IMC) \cite{Lyubartsev1995}, we find pair potentials that reproduce the quantum correlations of the free electron gas exactly. Now we can introduce these pair potentials as additional contributions to the Hamiltonian. 

A similar procedure was pursued originally by Lado \cite{Lado1967}, and subsequently used in the classical-map method of Perrot and Dharma-wardana for numerous applications \cite{Dharma-wardana2000,Dharma-wardana2008,Dharma-Wardana2012}. This technique has also proven to be successful in related semi-classical correlation studies \cite{Dufty2013,Dutta2013,Dutta2013b} and can be seen as analogous to correcting for the Pauli pressure in quantum hydrodynamics \cite{Manfredi2005}.

By insisting that the total potential is equal to the IMC result
$V_{\rm IMC}$, when applied to a non-interacting electron gas, we arrive
at an expression for the additional Pauli potential
\begin{equation}
V_{P}({\bf x})=-V_{B}({\bf x}|g_0)+\sum_{i,j}V_{\rm IMC}({\bf x}_{i},{\bf x}_{j}|g_0) \,.
\label{eq:PauliPotential}
\end{equation}
Here, $g_{0}$ is the ideal electron gas pair correlation, and $V_{B}(x|g_{0})$ is the Bohm potential evaluated at $g_{0}$. The full system potential $V$ is now given by
\begin{equation}
V=V_{ext}+V_{int}+V_{B}+V_{P}, \label{eq:TotalPotentialWithPauli}
\end{equation}
where $V_{ext}$ is an external potential (which is set to zero throughout our calculations), $V_{int}$ is the contribution from direct pair interactions of the simulated particles (e.g., the Coulomb potential), $V_{B}$ is the Bohm potential, and $V_{P}$ is the Pauli potential of Eq.~(\ref{eq:PauliPotential}).

\subsection*{Pseudopotentials}
In the description above, all particles are treated on equal footing. However, the core electrons of the ion species require an explicitly separate treatment from the (effectively free) valence electrons. The pseudopotential approximation, commonly used in implementations of DFT, provides a simple alternative to modeling the core electrons directly. Core electrons are removed from the system to be simulated and ion-electron potentials are constructed in a way that valence electrons form the correct density profile in a reference calculation. For our purposes, we use a reference single ion DFT calculation to determine an electron density profile and construct the pseudopotential. 

We use a variant of the Troullier-Martins approach also used for DFT pseudopotentials \cite{Troullier1991}. We first perform scalar-relativistic electron density calculations within DFT for an isolated ion with both $n_{core}$ and $n_{core}+1$ electrons. The difference of these densities is used as an input valence electron density. We then construct the pseudopotential by requiring the following properties:

\begin{enumerate}
\item The radially integrated electron density up to a cutoff, $r_{c}$, is the same for both the full ion system and pseudosystem. When solving for the two-body electron-ion pseudosystem via the Schr\"{o}dinger equation, we must have the same total electron density within $r_{c}$ as we found with the full DFT calculation. 
\item The pseudopotential for $r>r_{c}$ is equal to the (screened) Coulomb potential 
\item The pseudopotential is smooth and continuous at $r_{c}$ (Ref.~\cite{Troullier1991} discusses the specifics of the smoothing requirement)\end{enumerate}

Due to the direct connection to the approach used in standard DFT implementations, it should be possible to re-purpose existing pseudopotential generation codes for the method explained here. This should prove useful for materials with complex bound state structures where more sophisticated pseudopotential construction schemes have been shown to provide greater accuracy.

Accompanying the application of pseudopotentials is, however, the sacrifice of correctly treating bound-free electron transitions. As the number of valence electrons must be known a priori, as in implementations of orbital free DFT (OFDFT),
we must restrict our method here to cases where these numbers
are known and reasonably well-defined.

\subsection*{Generalized IMC Parameter Search \label{appendix:GeneralizedIMC}}
We now set out how the search for the optimal value of $\lambda$, the parameter in the linear Bohm potential, can be performed numerically. Due to the low computational demands associated with calculating $g$ for a given $\lambda$, we note that a brute force search should be adequate in most cases. However, a more robust approach, that is also applicable to potential forms with arbitrary numbers of free parameters, is possible through a generalization of IMC. We begin along the original lines of IMC by writing the positional Hamiltonian of the system as a sum over pair energies:
\begin{equation}
H=\sum_{\alpha}S_{\alpha}K_{\alpha}.
\end{equation}
Here, $K_{\alpha}$ represents the potential between two bodies at
a particular distance with discrete distance bins labelled by $\alpha$, and $S_{\alpha}$ is the number of pairs of bodies separated by that distance in the system, i.e., the number of pairs in the $\alpha^{th}$ bin. Our goal is to obtain a set of updates to a set of parameters defining the potential, $\Delta\lambda_{i}$, that we can effectively adjust the mean bin counts to match our desired pair correlations.

In our case, unlike the original
formulation of IMC, the potential $K_{\alpha}$ are functionals of the thermally averaged correlations of
the system, and are also functions of the chosen $\lambda$ values, via $K_{\alpha}=K_{\alpha}(\left\{ \left\langle S_{\alpha}\right\rangle \right\} ,\left\{ \lambda_{i}\right\} )$. Here the thermal expectation $\left\langle X\right\rangle$ of an arbitrary function of the system coordinates $X(q)$ is defined by
\begin{equation}
\left\langle X\right\rangle =\frac{1}{Z}\int X(q)\exp(-\beta H)dq \,,
\label{eq:IMCThermalAverage}
\end{equation}
where $Z$ is the partition function, $\beta=1/k_B T$, and $q$ is the set of all positional coordinates of the system. That is to say, the potentials of the system depend on both the set of $\lambda_i$ linearization parameters, and also the thermally averaged set of pair correlations between bodies (as per Eq.~\ref{eq:BohmPotential1d}).

Applying Eq.~(\ref{eq:IMCThermalAverage}) to $S_{\alpha}$ and differentiating with respect to $\lambda_{i}$, we obtain
\begin{equation}
\frac{\partial\left\langle S_{\alpha}\right\rangle }{\partial\lambda_{i}}=-\beta\sum_{\gamma}M_{i\gamma}(\left\langle S_{\alpha}S_{\gamma}\right\rangle -\left\langle S_{\alpha}\right\rangle \left\langle S_{\gamma}\right\rangle )\,, 
\label{eq:SAlphaDeriv}
\end{equation}
where the matrix $M_{i\gamma}$ is given by
\begin{equation}
M_{i\gamma}=\frac{\partial K_{\gamma}}{\partial\lambda_{i}}+\sum_{\delta}\frac{\partial K_{\gamma}}{\partial\left\langle S_{\delta}\right\rangle }\frac{\partial\left\langle S_{\delta}\right\rangle }{\partial\lambda_{i}} \,.
\label{eq:MMatrix}
\end{equation}
Combining Eqs.~(\ref{eq:SAlphaDeriv}) and (\ref{eq:MMatrix}), we find a separate set of linear equations for each $\lambda_i$
\begin{equation}
M_{i\gamma}=\frac{\partial K_{\gamma}}{\partial\lambda_{i}}-\beta\sum_{\epsilon}M_{i\epsilon}\sum_{\delta}\frac{\partial K_{\gamma}}{\partial\left\langle S_{\delta}\right\rangle }\left(\left\langle S_{\delta}S_{\epsilon}\right\rangle -\left\langle S_{\delta}\right\rangle \left\langle S_{\epsilon}\right\rangle \right).
\label{eq:MLinearEquations}
\end{equation}
Using these relationships, we can now perform a directed search for the optimal $\lambda_i$ values. To leading order, we can write an equation for changes in $\left\langle S_{\alpha}\right\rangle$ in terms of changes in ${\lambda_{i}}$
\begin{equation}
\Delta\left\langle S_{\alpha}\right\rangle =\sum_{i}\frac{\partial\left\langle S_{\alpha}\right\rangle }{\partial\lambda_{i}}\Delta\lambda_{i}.
\label{eq:DeltaSAlpha}
\end{equation}
If we perform a calculation of $g$ with a given set of $\lambda_i$ values, we can simultaneously calculate values of $\left\langle S_{\alpha}\right\rangle$ and $\left\langle S_{\alpha}S_{\gamma}\right\rangle$ numerically. We can then solve the linear equations (\ref{eq:MLinearEquations}) and, in turn, have an overdetermined set of
equations (Eq.~\ref{eq:DeltaSAlpha}) for the required changes in the $\lambda_i$ parameters to match the desired pair correlations. The optimal changes in $\lambda_i$
can then be determined by least-squares inversion of this equation.
Defining the matrix of values $A_{\alpha i}=\partial\left\langle S_{\alpha}\right\rangle /\partial\lambda_{i}$,
the formal solution is
\begin{equation}
\Delta\lambda=\left(A^{T}A\right)^{-1}A^{T}\Delta\left\langle S\right\rangle \,. 
\label{eq:delta_lambda}
\end{equation}
This can be determined directly, or implicitly through QR factorization \cite{Golub1996}
for computational efficiency. This update procedure can then be applied iteratively to arrive at the optimal parameters $\lambda_i$.

\subsection*{Modified Thermostats} \label{sec:Modified_Thermostats_Appendix}

Most modern DFT-MD simulations rely on either the Nos\'e-Hoover or Langevin thermostat to establish an ion dynamics at a given temperature \cite{White2013,Mabey2017,Witte2017}. In the case of the  Nos\'e-Hoover thermostat, an additional dynamic variable, coupled linearly to particle momenta through a friction term, is introduced to the equations of motion \cite{Nose1984}. The parameters for the thermostat can be chosen to ensure a balance between temperature stability and the equilibration time, but in general the dynamics are parameter-insensitive.

The Langevin thermostat, on the other hand, adds both a frictional term and a stochastic noise term to the equations of motion \cite{BinggeliN.Chelikowsky1994}. Again, this ensures the canonical distribution is sampled. The magnitude of the friction and noise are controlled by a free parameter, the Langevin friction $\sigma$, which must be chosen to be sufficiently small to minimize spurious effects on the particle dynamics, while remaining large enough that the correct distribution is sampled in a reasonable time-frame. An advantage of this approach is that $\sigma$ can be scaled up to a larger value to attempt to mimic electron-ion collisions, while still maintaining the canonical distribution \cite{Mabey2017}. In principle, the Born-Oppenheimer approximation can then be employed while still attempting to include the effects of the electron dynamics. However, the accurate setting of $\sigma$ for this purpose is a difficult task without an obvious solution (Ref.~\cite{Mabey2017} suggests some possible approaches, although the most appropriate choice is still unknown). Furthermore, this technique implicitly neglects all frequency dependence of the coupling between the electron and ion dynamics as the Langevin coupling is being approximated as white noise.

As our method does not rely on the Born-Oppenheimer approximation, we can employ the parameter-insensitive Nos\'e-Hoover thermostat for the ions. This represents the key advantage of our approach: the electron dynamics is modeled directly. Therefore, the ion thermostat -- and accordingly the ion dynamics we wish to calculate -- does not rely on the unknown free parameter $\sigma$ to mimic the dynamic electrons.

In dealing directly with dynamic electron trajectories, in contrast to Born-Oppenheimer DFT-MD, we must develop an appropriate thermostat for the electrons as well. In the degenerate and semi-degenerate cases, the velocity distribution of electrons is, however, known to be far from the Boltzmann distribution. While a standard thermostat allows us to move past the Born-Oppenheimer approximation and include dynamic electrons, it cannot capture the effect of Fermi-statistics on the electron motion. To more accurately capture the effect of the electron dynamics on the ions, we desire a thermostat that ensures static correlations are maintained, while enforcing a non-Boltzmann, that is Fermi, distribution of velocities.

We highlight here modifications of standard Langevin and Nos\'e-Hoover thermostats to achieve this aim.

\subsubsection*{Langevin-Style Thermostat}

The equations of motion of a particle in a stochastic force field can be written quite generally as:
\begin{equation}
d\boldsymbol{X}=\boldsymbol{\mu}dt+\boldsymbol{\sigma}\cdot d\boldsymbol{W},
\label{eq:LangevinGeneral}
\end{equation}
where 
\begin{equation}
\boldsymbol{X}=\left(X_{1},X_{2},X_{3},V_{1},V_{2},V_{3}\right)^{T},
\end{equation}
\begin{equation}
\boldsymbol{\mu}=(V_{1},V_{2},V_{3},\mu_{1}(\boldsymbol{X}),\mu_{2}(\boldsymbol{X}),\mu_{3}(\boldsymbol{X}))^{T}.
\end{equation}
Here, $\mu_{i}$ is the $i^{th}$ component of the deterministic part of the particle's acceleration. $\boldsymbol{X}$ is a vector containing the particle's position $(X_{1},X_{2},X_{3})^{T}$ and velocity $(V_{1},V_{2},V_{3})^{T}$, $\boldsymbol{\sigma}$ represents the stochastic collision frequency, and $\boldsymbol{W}$ is a multidimensional Wiener process \cite{Pavliotis2014}.
We take the driving noise to be uniform and isotropic in velocity
space. Hence,
\begin{equation}
\boldsymbol{\sigma}=\left(\begin{array}{ccc}
0 & 0 & 0\\
0 & 0 & 0\\
0 & 0 & 0\\
\sigma & 0 & 0\\
0 & \sigma & 0\\
0 & 0 & \sigma
\end{array}\right) \,.
\end{equation}
Defining
\begin{eqnarray}
\boldsymbol{V} & = & \left(V_{1},V_{2},V_{3}\right)^{T},\\
\boldsymbol{R} & = & (X_{1},X_{2},X_{3})^{T},\\
\boldsymbol{\eta} & = & (\mu_{1}(\boldsymbol{X}),\mu_{2}(\boldsymbol{X}),\mu_{3}(\boldsymbol{X}))^{T} \,,
\end{eqnarray}
we can then write, for this system, a Fokker-Planck equation for the equilibrium particle's probability density, $p$, in configuration space at long times \cite{Pavliotis2014}. We have
\begin{equation}
\frac{\partial}{\partial\boldsymbol{X}}\cdot(\boldsymbol{\mu}p)+\frac{\sigma^{2}}{2}\left(\frac{\partial}{\partial\boldsymbol{V}}\right)^{\!2}\!p=0 \,,
\end{equation}
which reduces to
\begin{equation}
-\boldsymbol{V}\cdot\frac{\partial p}{\partial\boldsymbol{R}}-\frac{\partial}{\partial\boldsymbol{V}}\cdot(\boldsymbol{\eta}p)+\frac{\sigma^{2}}{2}\left(\frac{\partial}{\partial\boldsymbol{V}}\right)^{\!2}\!p=0 \,.
\label{eq:FokkerPlanck}
\end{equation}
We approximate the probability density distribution of the particles as decoupled in momentum and position
space, via
\begin{equation}
p\equiv p_L=A\frac{1}{1+e^{\beta(m\boldsymbol{V}^{2}/2-\mu_{c})}}\cdot e^{-\beta U(\boldsymbol{R})},
\label{eq:PLDefinition}
\end{equation}
where $U$ is the potential energy, $A$ is a normalization constant, and $\mu_{c}$ is the chemical potential. By inserting $p$ from eq. (\ref{eq:PLDefinition}) into eq. (\ref{eq:FokkerPlanck})
we can solve for $\boldsymbol{\eta}$; the equations of motion in eq. (\ref{eq:LangevinGeneral}) can then be written as 
\begin{equation}
d\boldsymbol{R}=\boldsymbol{V}dt\label{eq:r_dot}
\end{equation}
\begin{multline}
d\boldsymbol{V}=\frac{\boldsymbol{F}}{m}(1+e^{\beta E})\left[\log(1+e^{\beta E})-\beta E\right]dt\\
-\frac{\sigma^{2}}{2}m\beta\left(\frac{e^{\beta E}}{1+e^{\beta E}}\right)\boldsymbol{V}dt+\boldsymbol{\sigma}\cdot d\boldsymbol{W} \,,
\label{eq:v_dot}
\end{multline}
where $m$ is the particle's mass, $E=m\boldsymbol{V}^{2}/2-\mu_{c}$ and $\boldsymbol{F}=-\partial U/\partial\boldsymbol{R}$. The boundary conditions were set to ensure that the standard Langevin equation is recovered at high energy. These represent the modified equations of motion, solvable with standard methods, which have the desired equilibrium distribution function.

\begin{figure}
\centering{}\includegraphics[width=0.95\columnwidth]{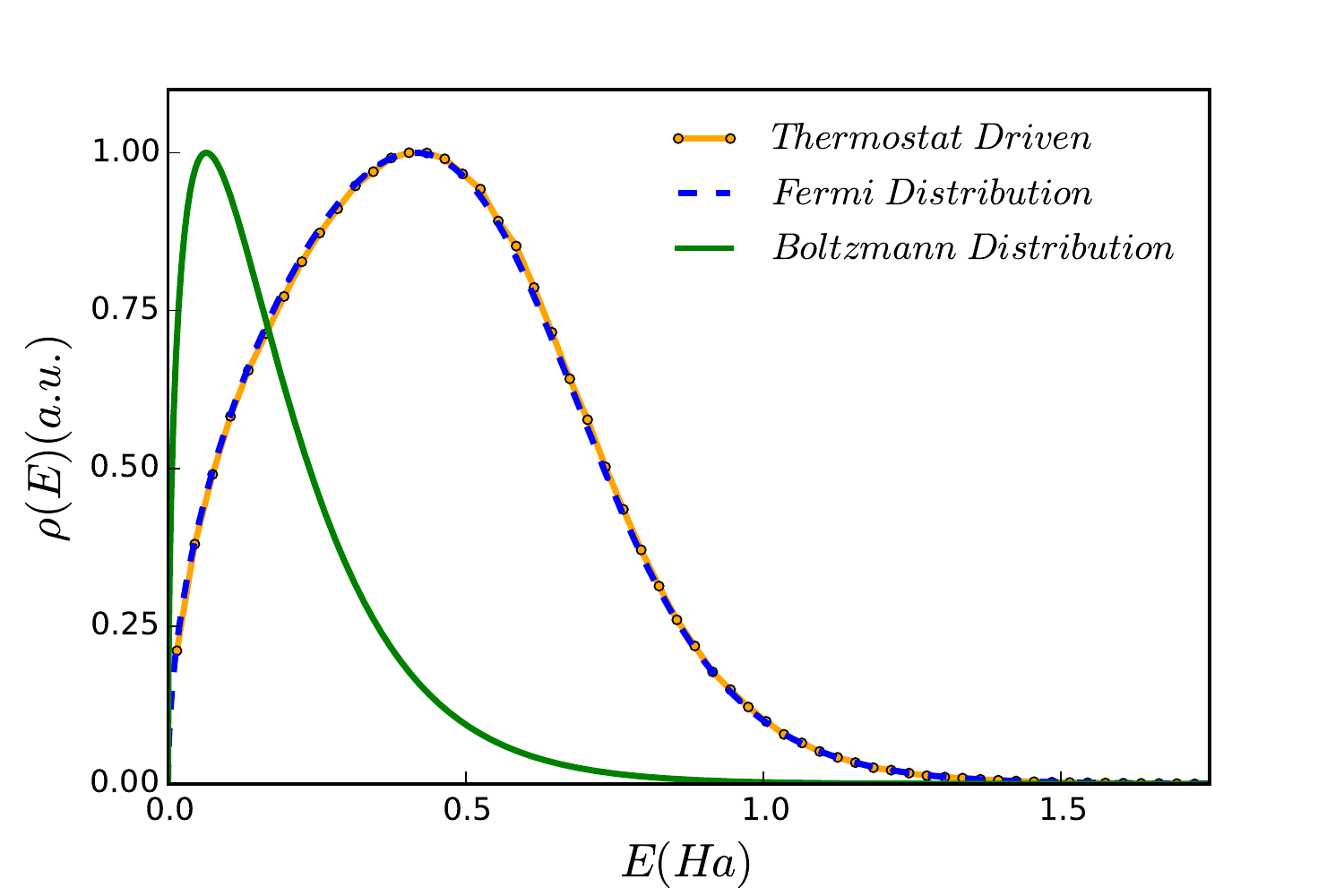}\caption{Reproduction of a Fermi kinetic energy distribution using our modified thermostat. The system considered was comprised of particles at the same density/temperature as valence electrons in twice-compressed-Al at $3.5$ eV. A simple exponential pair potential $V=V_{0}\exp(-\kappa r)$ with $V_{0}=1$ Ha and $\kappa=1$ $a_{B}^{-1}$ was employed. Distributions are normalized to have maxima of one.\label{fig:thermostat_example}}
\end{figure}

\subsubsection*{Nos\'e-Hoover-Style Thermostat}

By analogy with the Langevin case, we begin with the equations for the dynamics
\begin{equation}
\boldsymbol{\dot{R}}=\boldsymbol{V} \,,
\label{eq:r_dot-1}
\end{equation}
\begin{eqnarray}
\boldsymbol{\dot{V}}=\frac{\boldsymbol{F}}{m}(1+e^{\beta E})\left[\log(1+e^{\beta E})-\beta E\right] \nonumber \\
- \frac{\alpha}{2}m\beta\left(\frac{e^{\beta E}}{1+e^{\beta E}}\right)\boldsymbol{V}\,,
\label{eq:v_dot-1}
\end{eqnarray}
where $\alpha$ is a new dynamic variable coupling the system to the heat bath. In equilibrium, our deterministic system must now satisfy the generalized Liouville equation \cite{Martyna1992}
\begin{multline}
\frac{\partial p_{NH}}{\partial\boldsymbol{R}}\cdot\boldsymbol{\dot{R}}+\frac{\partial p_{NH}}{\partial\boldsymbol{V}}\cdot\boldsymbol{\dot{V}}+\frac{\partial p_{NH}}{\partial\alpha}\dot{\alpha}\\
+p_{NH}\left(\frac{\partial}{\partial\boldsymbol{R}}\cdot\boldsymbol{\dot{R}}+\frac{\partial}{\partial\boldsymbol{V}}\cdot\boldsymbol{\dot{V}}+\frac{\partial\dot{\alpha}}{\partial\alpha}\right)=0 \,,
\end{multline}
where the probability distribution $p_{NH}$ is given by $p_{NH}=p_{L}\exp(-\beta\alpha^{2}/2m_{\alpha})$ -- in which we force $\alpha$ to have a Gaussian distribution, in analogy with the standard Nos\'e-Hoover scheme -- and $m_{\alpha}$
is a thermostat mass. Again, asserting that we must recover the standard
Nos\'e-Hoover thermostat in the classical limit, we can solve this equation for $\dot{\alpha}$. We obtain
\begin{equation}
\frac{\dot{\alpha}}{m_{\alpha}}=-\frac{m}{2}\left(\frac{e^{\beta E}}{1+e^{\beta E}}\right)\left[\beta m\boldsymbol{V}^{2}\left(\frac{1-e^{\beta E}}{1+e^{\beta E}}\right)+3\right] \,,
\end{equation}
which, in combination with Eqs.~(\ref{eq:r_dot-1}) and (\ref{eq:v_dot-1}), forms a closed dynamical system with the desired equilibrium distribution. For multiple particles, $\dot{\alpha}$ is just a sum over terms of
this form for each particle. Fig.~\ref{fig:thermostat_example} demonstrates the recovery of the desired momentum distribution for a sample system.

\subsection*{Simulation Parameters}

The linearization factors and Bohm potentials were constructed according to the theory described in the main paper. The information about static correlations needed as input, specifically the radial distribution function of the ionic system, was produced with OFDFT calculation running the using VASP package \cite{Kresse1993,Kresse1994,Kresse1996,Kresse1996a} for short times. For this purpose, we used plane wave and augmentation energies of $250$ eV and $500$ eV, respectively, the 3e Al pseudopotential provided with VASP, and a simulation box containing 256 atoms. A total of $2680$ electron bands were included, such that the occupations of the highest energy bands were less than $10^{-6}$. This simulation was run for $5000$ time steps of $1$\,fs.

Electron density profiles required to produce pseudo\-potentials were calculated using the GPAW DFT package \cite{Enkovaara2010,Mortensen2005,Larsen2017} with the PBE functional \cite{Perdew1996}. The MD simulation step was carried out with a standard MD code, inclusive of our novel thermostat. As usual, periodic boundary conditions were employed. The simulation box contained 256 aluminum ions and the simulations were run with a time step of $0.1\,\,\rm a.u.$ All properties were extracted by averaging over multiple simulation runs each having $10^{7}$ time steps.

\bibliographystyle{unsrt}